\documentclass[9pt,twocolumn,twoside]{article}
\usepackage[utf8]{inputenc}
\usepackage{fullpage}
\usepackage[top=2cm, bottom=2cm, left=1.2cm, right=1.2cm]{geometry}
\usepackage[english]{babel}
\usepackage{amsmath}
\usepackage{hyperref}

\usepackage{authblk}
\usepackage{graphicx}
\usepackage{bm}

\newcommand{\micron}{\textmu}

\title{Efficient Type II Second Harmonic Generation in an Indium Gallium Phosphide on insulator wire waveguide aligned with a crystallographic axis}

\author[*,1,2,3]{Nicolas Poulvellarie}
\author[1]{Carlos Mas Arabi}
\author[4]{Charles Ciret}
\author[5]{Sylvain Combrié}
\author[5]{Alfredo De Rossi}
\author[1]{Marc Haelterman}
\author[6,7]{Fabrice Raineri}
\author[2,3]{Bart Kuyken}
\author[1]{Simon-Pierre Gorza}
\author[1]{Fran\c{c}ois Leo}

\affil[1]{OPERA-Photonique, Universit\'{e} libre de Bruxelles, Brussels, Belgium}
\affil[2]{Photonics Research Group, Ghent University-IMEC, Ghent, Belgium}
\affil[3]{Center for Nano and Biophotonics (NB-Photonics), Ghent University-IMEC, Ghent, Belgium}
\affil[4]{Laboratoire de Photonique d'Angers EA 4464, Université d'Angers, Angers, France}
\affil[5]{Thal\`{e}s Research and Technology, Palaiseau, France}
\affil[6]{Centre de Nanosciences et de Nanotechnologies (C2N), CNRS, Universit\'{e} Paris Sud, Universit\'{e} Paris Saclay, F-91120 Palaiseau, France}
\affil[7]{Universit\'{e} de Paris, Centre de Nanosciences et de Nanotechnologies (C2N), F-91120 Palaiseau, France}

\affil[*]{Nicolas Poulvellarie: Nicolas.Poulvellarie@ulb.be}

\begin{document}

\twocolumn[
\begin{@twocolumnfalse}
\maketitle
\begin{abstract}
We theoretically and experimentally investigate type II second harmonic generation in III-V-on-insulator wire waveguides. We show that the propagation direction plays a crucial role and that longitudinal field components can be leveraged for robust and efficient conversion.
We predict that the maximum theoretical conversion is larger than that of type I second harmonic generation for similar waveguide dimensions and reach an experimental conversion efficiency of 12~\%/W, limited by the propagation loss.
\end{abstract}
\bigskip{}
\end{@twocolumnfalse}
]

Second harmonic generation (SHG) was first demonstrated almost 60 years ago \cite{franken_generation_1961} and still attracts a lot of attention.
In particular, many novel integrated platforms for integrated SHG have recently emerged \cite{schneider_gallium_2018, levy_harmonic_2011, logan_400w_2018}. Conversions as high as 47000~\%/(W.cm$^2$) have been demonstrated in III-V semiconductors \cite{stanton_efficient_2020} and ultra-wide tuning was recently shown in silicon waveguides \cite{singh_broadband_2020}. The large intrinsic nonlinearity in addition to the high-index contrast of sub-wavelength structures leads to very large effective nonlinearities.
Different schemes for phase matching have been proposed and implemented~\cite{bartnick_cryogenic_2020, wang_ultrahigh-efficiency_2018, nitiss_formation_2020}. Type I modal phase matching, consisting in engineering the waveguide cross section such that a pump mode and a SH higher-order mode propagate at the same phase velocity, is attractive because standard waveguiding structures can be used.
It however leads to lower conversion efficiency as compared to other schemes because of the sub-optimal overlap between the pump and SH mode in the nonlinear core of the waveguide.
Record conversions in gallium arsenide wire waveguides were recently reported by optimizing such overlap. 
But the efficiency is still limited because very thin layers are used to ensure phase matching between two fundamental modes~\cite{stanton_efficient_2020}. Moreover, it makes the conversion very sensitive to fabrication variations.

Here we show that type II SHG can help alleviate these limitations. We theoretically analyze the nonlinear conversion with a full-vectorial model to identify the most efficient configurations and experimentally confirm our predictions in indium gallium phosphide (InGaP) nanowires.

To the best of our knowledge, type II SHG has never been demonstrated in the strong-guidance regime. 
It was previously studied in AlGaAs photonic wires with low vertical confinement~\cite{duchesne_second_2011}, where the modes are well approximated by transverse waves. As we recently demonstrated for the case of type I SHG, more complex wave mixing involving the longitudinal components can be expected in III-V-on-insulator wire waveguides~\cite{poulvellarie_second-harmonic_2020, ciret_influence_2020}. We hence apply the same vectorial analysis to the case of type II phase matching in order to identify efficient nonlinear couplings between a pump around 1550~nm and a higher order mode around 775~nm.

\begin{figure}[ht] 
\centering
\includegraphics[scale=1]{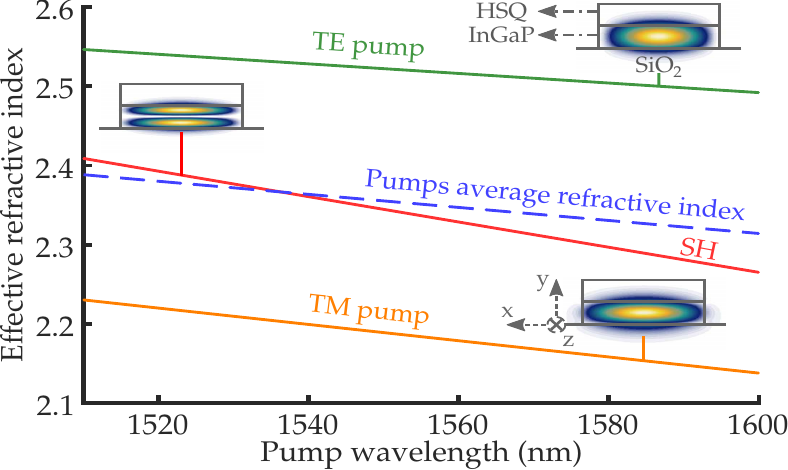} 
\caption{Effective refractive indices of the pump and SH modes of an InGaP-on-insulator wire waveguide as a function of the pump wavelength. The corresponding Poynting vector distributions are also shown.
The average of the indices of the two pump fundamental modes is plotted to highlight the phase matching point. }
\label{PM}
\end{figure}

We write the electric field as a superposition of three forward propagating bound modes, two oscillating at $\omega_0$ and one at $2\omega_0$, in the waveguide frame ($xyz$), where $x$ is the horizontal coordinate, $y$ the vertical coordinate and $z$ the propagation direction (see Fig.~\ref{PM}). The total electric field reads:

\begin{multline}
\bm{E} = a_{1}(z)\bm{e}_{a_1}(\mathbf{r}_\perp,\omega_0) e^{i(\beta_{a_1}z - \omega_{0}t)} + a_{2}(z)\bm{e}_{a_2}(\mathbf{r}_\perp,\omega_0) e^{i(\beta_{a_2}z - \omega_{0}t)}\\
\qquad + b(z)\bm{e}_{b}(\mathbf{r}_\perp,2\omega_0) e^{i(\beta_{b}z - 2\omega_{0}t)} + c.c.,
\label{eqn:ElectricField}
\end{multline}

where $a_{1}$ and $a_{2}$ represent the amplitudes of both pump modes and $b$ the amplitude of the SH mode (expressed in $\sqrt{W}$). $\beta_{a_1}$, $\beta_{a_2}$ are the propagation constants at the carrier frequency $\omega_0$ and $\beta_{b}$ is the propagation constant at the carrier frequency 2$\omega_0$. $\bm{e}_{a1,~a2}(x,y,\omega_{0})$ and $\bm{e}_{b}(x,y,2\omega_{0})$ are the orthonormal vectorial electric profiles of the modes. 
They satisfy the usual orthonormality condition~\cite{snyder_optical_1983}.

The propagation equations can be found through perturbation theory \cite{afshar_v_full_2009, alloatti_second-order_2012, kolesik_nonlinear_2004} .
Here, we only retain the nonlinear terms involving three different modes and include the propagation loss. We find:

\begin{align}
\begin{split}
\frac{d a_{1}}{dz} &= - \frac{\alpha_{a_1}}{2}a_{1} + i\kappa_{12}^*ba_{2}^*\exp{(-i\Delta\beta z)} \\
\frac{da_{2}}{dz} &= - \frac{\alpha_{a_2}}{2}a_{2} + i\kappa_{12}^*ba_{1}^*\exp{(-i\Delta\beta z)} \\
\frac{db}{dz}   &= - \frac{\alpha_{b}}{2}b + 2i\kappa_{12}a_{1}a_{2}\exp{(i\Delta\beta z)} \\
\end{split}
\label{eqn:Coupled}
\end{align}

where $\alpha_{a_1}$, $\alpha_{a_2}$ and $\alpha_{b}$ are the linear loss coefficients of the respective mode and $\Delta\beta = \beta_{a1} + \beta_{a2} - \beta_{b}$ is the wavenumber mismatch.
The effective nonlinearity $\kappa_{12}$, expressed in $\mathrm{(\sqrt{W}.m)^{-1}}$, reads

\begin{align}
\kappa_{12} = \omega_0 \varepsilon_0 \iint_A \sum_{jkl=xyz}\chi_{jkl}^{(2)}e_{b}^{*j}e_{a_1}^{k}e_{a_2}^{l}dxdy,
\label{eqn:kappa}
\end{align}

with $\varepsilon_0$, the vacuum permittivity and $A$, the nonlinear waveguide core.
We recall that Eqs.(\ref{eqn:Coupled}) are written in the waveguide frame such that the nonlinear tensor elements are dependent on the propagation direction. In the crystal frame of a III-V semiconductor, only the $j \neq k\neq l$ elements are nonzero. To account for other directions, the nonlinear tensor must be rotated (see e.g.~\cite{duchesne_second_2011}).
Most III-V wafers are grown along a crystal axis. Here we consider the case of a wafer grown along the [010] crystallographic axis and use the [100] direction as the reference in the propagation plane.  

In the undepleted regime, the SH output power can be exactly calculated. We neglect the parametric down conversion terms and integrate \eqref{eqn:Coupled} along the propagation direction. We find

\begin{align}
P_{sh}&=16\left|\kappa_{12}\right|^{2}\frac{P_{1}P_{2}}{\left|\Phi\right|^{2}}\left|\sinh{\left(\frac{\Phi}{2}z\right)}\right|^{2}\exp{\left[-\left(\frac{\alpha_{b}}{2}+\alpha_a\right)z\right]},
\label{eqn:ShPower}
\end{align}

where $\alpha_{a_1} = \alpha_{a_2} = \alpha_a$ and $ \Phi = \alpha_{b}/{2}-\alpha_a+i\Delta\beta$. $P_{1,2}=\left|E_{1,2}\right|^2$ are the input powers at the pump wavelength. To maximize the conversion, we set $P_{1}=P_{2}=P_{0}/2$ with $P_0$ the total input power. The maximum output power, found by setting $\Delta\beta=0$, can then be written $P_{sh}(L) = \left|\kappa P_0 L_{\mathrm{eff}}\right|^2$, where

\begin{align}
L_\mathrm{eff} &= 2 \frac{\exp{\left(-\alpha_{a}L\right)} - \exp{\left(-\alpha_{b}L/{2}\right)}}{\alpha_{b} - 2\alpha_{a}},
\label{eqn:Leff}
\end{align}

and L the length of the waveguide. This is the same expression as that of type I SHG~\cite{poulvellarie_second-harmonic_2020}. In what follows we use the effective nonlinearity $\kappa_{12}$ to compare different configurations as well as to confront our experimental results with theoretical predictions.

\begin{figure}[h] 
\centering
\includegraphics[scale=1]{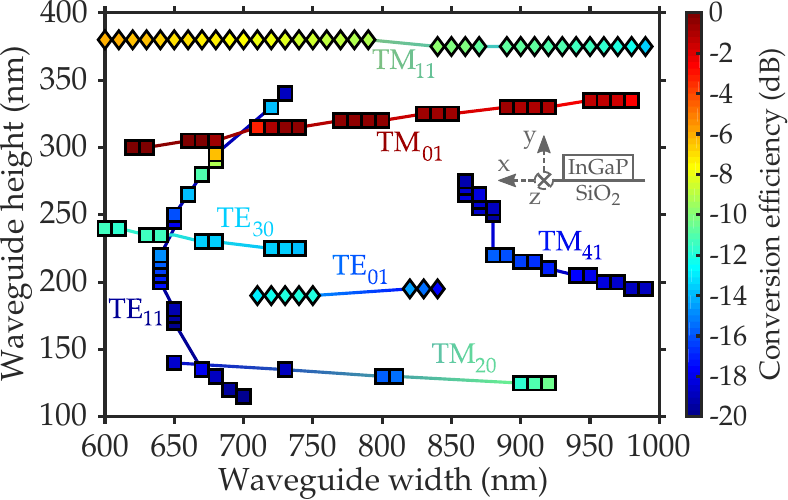} 
\caption{Efficiency map of the nonlinear coupling between $\mathrm{TE_{00}}$ and $\mathrm{TM_{00}}$ pump modes and a higher order second harmonic mode of an air-clad InGaP-on-insulator wire. Only phase-matched interactions are shown.
Squares (diamonds) correspond to a waveguide rotated at 0$^{\circ}$ (45$^{\circ}$) from the [100] crystal axis. The conversion efficiency is normalized to the maximum value [$\left(|\kappa|/|\kappa_{max}|\right)^2$ (dB) with $\kappa_{max} = i3675~\mathrm{(W^{1/2}.m)^{-1}}$].}
\label{Map}
\end{figure}

We now look for specific cases of phase matching to evaluate the theoretical conversion efficiency in standard waveguides. We first consider air-clad, fully-etched, InGaP-on-insulator wire waveguides. The single nonzero second order nonlinear coefficient was measured to be as high as $\chi^{(2)}=220$~pm/V~\cite{ueno_second-order_1997}.
In principle, any two pump modes can be used for type II SHG, but wave mixing processes involving the two fundamental modes (TE$_{00}$ and TM$_{00}$) are expected to be the most efficient. Moreover, it is the easiest scheme to implement experimentally

in sub-wavelength waveguides with free-space injection.
We hence look for phase matching between the two fundamental modes around 1550~nm and a higher-order mode around 775~nm. We vary the width and height of the waveguide in steps of 5~nm in each direction. When a phase matching point is found in a 10~nm window around the 1550~nm wavelength, we compute the efficiency as a function of the propagation direction (either $0^\circ$ or $45^\circ$) and place a marker in the efficiency map shown in Fig.~\ref{Map}. 

The marker color codes the relative strength of the coupling and its shape indicates the propagation direction for which it is maximized: squares correspond to 0$^{\circ}$ waveguides while diamonds represent 45$^{\circ}$ waveguides.
To limit the computational time, we restrict the simulations to widths between 600 and 1000~nm and heights between 100 and 400~nm. We connect the markers corresponding to the same SH higher-order mode with a line.
Interestingly, and in contrast to type I SHG~\cite{ciret_influence_2020}, the most efficient conversion occurs for relatively thick waveguides (620~nm wide, 300~nm high) aligned with a main crystallographic axis [$\kappa_{max} = i3675~\mathrm{(W^{1/2}.m)^{-1}}$]. 
For comparison,  the maximum found for type I SHG in the same range of waveguide dimensions is $\kappa_1 = 2816~\mathrm{(W^{1/2}.m)^{-1}}$ ~\cite{ciret_influence_2020}. 
The larger efficiency for type II SHG can be understood by analyzing
the expression of the effective nonlinearity for waves propagating along a crystallographic axis of a III-V material. It reads
\begin{multline}
\kappa_{12} = \omega_0 \varepsilon_0 \iint_A \chi_{xyz}^{(2)}[
 e_{b}^{*x}(e_{a_1}^{y}e_{a_2}^{z}+e_{a_1}^{z}e_{a_2}^{y}) + \\
 e_{b}^{*y}(e_{a_1}^{z}e_{a_2}^{x}+e_{a_1}^{x}e_{a_2}^{z})+ e_{b}^{*z}(e_{a_1}^{x}e_{a_2}^{y}+e_{a_1}^{y}e_{a_2}^{x})] dxdy,
 \label{eqn:kappaxyz}
\end{multline}

where $\chi^{(2)}_{xyz}$ is the single nonzero tensor element of InGaP. In the specific case of predominant TM$_{00}$ and TE$_{00}$ pump modes and a TM higher-order mode at the SH, \eqref{eqn:kappaxyz} can be approximated by the simpler expression:
\begin{equation}
 \kappa_{12}\approx\omega_0 \varepsilon_0 \iint_A \chi_{xyz}^{(2)}e_{a_1}^{x} (e_{b}^{*y}e_{a_2}^{z}+e_{b}^{*z}e_{a_2}^{y})dxdy.
 \label{eqn:kappaapprox} 
\end{equation}

The effective nonlinearity is dominated by two terms, both involving a longitudinal electric field component. A transverse mode approximation would yield no conversion in this case. The wave mixing process involving longitudinal components does not preclude efficient conversion because, in high index contrast platforms, the longitudinal components can be almost as large as their transverse counterpart~\cite{driscoll_large_2009}. Importantly, longitudinal components have a spatial distribution that is distinct from, but linked to, that of the principal transverse component~\cite{snyder_optical_1983}.
In the case of phase matching to a TM$_{01}$ mode, both terms of \eqref{eqn:kappaapprox} are large because the involved spatial distributions are very similar to each other. 
This is made possible by the very large index contrast of the platform and highlights the strong potential of type II phase matching for SHG in III-V-on-insulator wire waveguides. 

Next we aim to experimentally confirm this large theoretical conversion efficiency.

We fabricate InGaP on insulator waveguides through wafer bonding~\cite{dave_nonlinear_2015}. The starting epitaxial stack is made of a 350~\micron m substrate of AlGaAs, a 200~nm sacrificial layer of InGaP, another 200\,$\mathrm{nm}$ sacrificial layer of AlGaAs and then a 320~nm InGaP layer. A thin layer of 15~nm of silicon oxide is deposited on top of the stack to improve adhesion. The stack is then bonded on a oxidized silicon wafer (3~\micron m SiO$_2$) using a BenzoCycloButene (BCB) dilution as an adhesive layer. We remove the substrate with a $\mathrm{HNO_3:H_2O_2:H_2O}$ solution in a 1:4:1 proportion and pattern waveguides using electron-beam lithography.
A negative resist [hydrogen silsesquioxane (HSQ)] is deposited prior to illumination. The waveguides are patterned using ICP etching. The electron-beam resist is not removed resulting in a 200~nm cladding with an index similar to that of silicon dioxide (see Fig.~\ref{PM}).

\begin{figure}[ht] 
\centering
\includegraphics[scale=1]{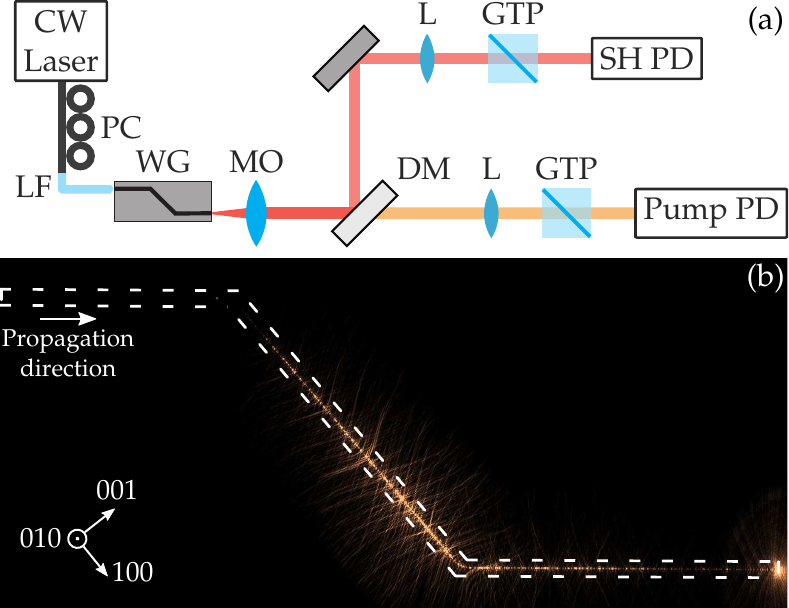}
\caption{(a) Experimental set-up. A continuous wave (CW) laser is injected in the waveguide (WG) thanks to a lensed fiber (LF). The light is collected with a microscope objective (MO) before being split by a dichroic mirror (DM). Each wave goes through an imaging lens (L) and a Glan-Taylor polarizer (GTP). The power is finally measured by a dedicated photodiode (PD).
(b) Top view of the waveguide as captured with a silicon camera.}
\label{Setup}
\end{figure}

The width of the waveguides, characterized with scanning electron microscopy, is 850~nm. In this structure, the efficient conversion to a TM$_{01}$ SH mode, as discussed above, is predicted to occur for a pump wavelength of 1536~nm (see Fig.~\ref{PM}). The corresponding theoretical effective nonlinearity is $\kappa_{12} = i3200~\mathrm{(\sqrt{W}.m)^{-1}}$.
We design waveguides made of three sections of different directions [see Fig.~\ref{Setup}(b)]. This is because the cleave direction of both silicon and III-V semiconductors are at 45$^{\circ}$ (i.e. along the [101] and [$\mathrm{\bar{1}01}$] axis).
The main section, located in the middle, is aligned with a crystal axis.
It is connected, on both ends, to sections normal to cleavage planes to facilitate light injection and collection. A 5~\micron m wide and 200~\micron m long taper is used at the input to optimize the injection. The middle section is 1.4~mm long and the total length is 4.5~mm.

The experimental setup is depicted in Fig.~\ref{Setup}(a). We inject the light using a lensed fiber. The coupling loss is estimated at 6~dB. A polarization controller allows us to tune the input polarization state. The light is collected from the waveguide with a high NA objective (0.9). The pump and second harmonic wavelengths are separated with a dichroic mirror and sent, through a Glan-Taylor polarizer, to a photodiode.
To characterize the SH diffusion pattern, we image the chip from the top with a silicon camera. In a first experiment, we inject 25~mW from a C-band laser and tune both the input wavelength and polarization. As predicted, the brightest pattern is found at 1536~nm when both a TE$_{00}$ and a TM$_{00}$ pump mode are simultaneously excited. The output SH wave is predominantly vertically polarized, as expected.
The corresponding SH diffusion pattern is shown in Fig.~\ref{Setup}(b). We see that the SH appears at the beginning of the middle section, the one aligned with a crystallographic axis. The SH intensity increases until the second bend, after which it decays, as expected from the lack of nonlinear coupling in waveguides rotated at 45$^{\circ}$ from a crystal axis.
To extract the experimental effective nonlinearity ($\kappa_{\mathrm{exp}}$), we estimate the loss at each wavelength by fitting diffusion patterns.
For the pump modes, we perform an experiment at low power and for the SH TM$_{01}$ mode we use the last section of the waveguide where no SH conversion occurs. We find 1.2~dB/mm for the pump modes and 6.4~dB/mm at the second harmonic, corresponding to $L_{eff} = 700$~\micron m. An injected input power of 25~mW corresponds to 3.8~mW at the beginning of the 0$^{\circ}$ section. Because it is difficult to experimentally evaluate the outcoupling loss for the SH mode, we consider that we collect all the output power and hence give a lower bound of the experimental effective nonlinearity.
We first extract $\kappa_{\mathrm{exp}}$ from the output SH power at the phase matching wavelength and next characterize the spectral transmittance of the process.

\begin{figure}[ht] 
\centering
\includegraphics[scale=1]{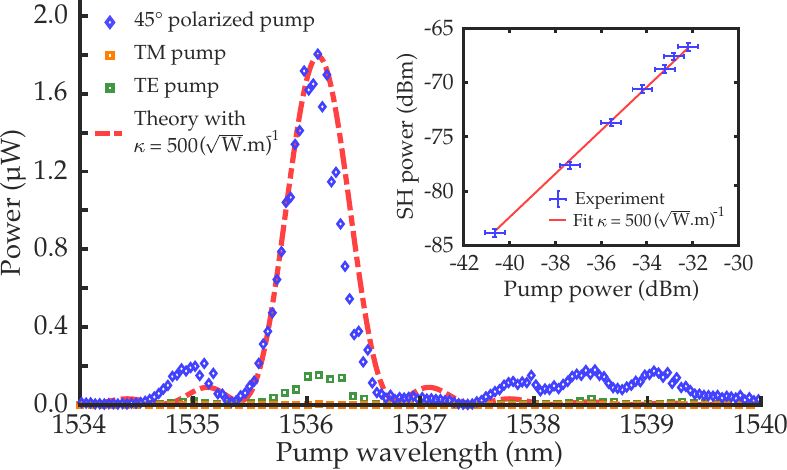} 
\caption{Experimental SH power as a function of the pump wavelength for a purely TE pump (green squares), a purely TM pump (orange squares) and a hybrid pump (blue diamonds). The red dashed line corresponds to the theoretical output SH power.
The inset shows the measured (blue crosses) and fit (red line) of the SH power as a function of the input power.}
\label{Sinc}
\end{figure}

The maximum output SH power (at $\lambda_{p} = 1536~\mathrm{nm})$ as a function of the input power is shown in the inset of Fig.~\ref{Sinc}. It is fitted with the theoretical function $P_{sh}=\left|\kappa_\mathrm{exp} P_0 L_{\mathrm{eff}}\right|^2$ that allows us to extract the conversion efficiency value.
We find $|\kappa_\mathrm{exp}| > 500~\mathrm{(\sqrt{W}.m)^{-1}}$.
The recorded output SH power as a function of input wavelength for three different input polarization states is shown in Fig.~\ref{Sinc} and confirms that the process is more efficient when both input modes are equally excited. Also plotted is the theoretical transfer function \eqref{eqn:ShPower} computed for $|\kappa_\mathrm{exp}|=500~\mathrm{(\sqrt{W}.m)^{-1}}$. The agreement between the theoretical spectral acceptance and our experimental results is excellent.

The maximum conversion efficiency, defined as $\left|\kappa_\mathrm{exp}\right|^2$, is equal to 2500~\%/(W.cm$^2$), which corresponds to 12~\%/W in our waveguide.
It is larger than recently reported values in lithium niobate [41~\%/(W.cm$^2$)]~\cite{wang_second_2017} and gallium phosphide [0.3~\%/(W.cm$^2$)]~\cite{anthur_second_2020}.
It is however an order of magnitude lower than the record 47000~\%/(W.cm$^2$) efficiency recently obtained in GaAs nanowaveguides~\cite{stanton_efficient_2020}. 
Yet, the theoretical limit in our case, corresponding to $|\kappa_{12}|^2$, is 102400~\%/(W.cm$^2$). It highlights the strong potential of type II SHG for future integrated frequency converters. InGaP wire waveguides with much lower propagation loss should hence permit converting as little as 1~\micron W into 1~nW of SH power in a 1~cm long waveguide. Encouragingly, several demonstrations of low loss III-V semiconductor waveguides have been recently reported~\cite{ottaviano_low-loss_2016, chang_ultra-efficient_2020}.

In conclusion, we demonstrated efficient type II SHG in InGaP nanowires. We performed a full-vectorial theoretical analysis and showed that the most efficient conversion occurs in waveguides aligned with a crystallographic axis. In that configuration, the nonlinear coupling is enabled by longitudinal field components of both a pump mode and the SH mode. We confirmed our prediction experimentally by demonstrating very efficient conversion in a 850~nm wide, 320~nm thick wire waveguide. As predicted, the conversion is maximized when the two fundamental modes at the pump wavelength are equally excited and the conversion occurs in a waveguide aligned with a main crystallographic axis. These results demonstrate the potential of type II phase matching to maximize the conversion in III-V semiconductor nanowaveguides, which we expect to play a role in future quantum circuits~\cite{zaske_visible--telecom_2012} and frequency comb stabilization devices~\cite{diddams_direct_2000, okawachi_-chip_2020}.

\medskip
\noindent\textbf{Disclosures.} The authors declare no conflicts of interest.

\bibliographystyle{unsrt}
\bibliography{Bib15.bib}

\begin{thebibliography}{10}

\bibitem{franken_generation_1961}
P.~A. Franken, A.~E. Hill, C.~W. Peters, and G.~Weinreich.
\newblock Generation of {Optical} {Harmonics}.
\newblock {\em Physical Review Letters}, 7(4):118--119, August 1961.

\bibitem{schneider_gallium_2018}
Katharina Schneider, Pol Welter, Yannick Baumgartner, Herwig Hahn, Lukas
  Czornomaz, and Paul Seidler.
\newblock Gallium {Phosphide}-on-{Silicon} {Dioxide} {Photonic} {Devices}.
\newblock {\em Journal of Lightwave Technology}, 36(14):2994--3002, July 2018.

\bibitem{levy_harmonic_2011}
Jacob~S. Levy, Mark~A. Foster, Alexander~L. Gaeta, and Michal Lipson.
\newblock Harmonic generation in silicon nitride ring resonators.
\newblock {\em Optics Express}, 19(12):11415, June 2011.

\bibitem{logan_400w_2018}
Alan~D. Logan, Michael Gould, Emma~R. Schmidgall, Karine Hestroffer, Zin Lin,
  Weiliang Jin, Arka Majumdar, Fariba Hatami, Alejandro~W. Rodriguez, and
  Kai-Mei~C. Fu.
\newblock 400\%/{W} second harmonic conversion efficiency in 14 um-diameter
  gallium phosphide-on-oxide resonators.
\newblock {\em Optics Express}, 26(26):33687, December 2018.

\bibitem{stanton_efficient_2020}
Eric~J. Stanton, Jeff Chiles, Nima Nader, Galan Moody, Nicolas Volet, Lin
  Chang, John~E. Bowers, Sae Woo~Nam, and Richard~P. Mirin.
\newblock Efficient second harmonic generation in nanophotonic
  {GaAs}-on-insulator waveguides.
\newblock {\em Optics Express}, 28(7):9521, March 2020.

\bibitem{singh_broadband_2020}
Neetesh Singh, Manan Raval, Alfonso Ruocco, and Michael~R. Watts.
\newblock Broadband 200-nm second-harmonic generation in silicon in the telecom
  band.
\newblock {\em Light: Science \& Applications}, 9(1):17, December 2020.

\bibitem{bartnick_cryogenic_2020}
Moritz Bartnick, Matteo Santandrea, Jan~Philipp Hoepker, Frederik Thiele,
  Raimund Ricken, Viktor Quiring, Christof Eigner, Harald Herrmann, Christine
  Silberhorn, and Tim~J. Bartley.
\newblock Cryogenic second harmonic generation in periodically-poled lithium
  niobate waveguides.
\newblock {\em arXiv:2005.07500 [physics, physics:quant-ph]}, May 2020.
\newblock arXiv: 2005.07500.

\bibitem{wang_ultrahigh-efficiency_2018}
Cheng Wang, Carsten Langrock, Alireza Marandi, Marc Jankowski, Mian Zhang,
  Boris Desiatov, Martin~M. Fejer, and Marko Loncar.
\newblock Ultrahigh-efficiency second-harmonic generation in nanophotonic
  {PPLN} waveguides.
\newblock {\em Optica}, 5(11):1438, November 2018.
\newblock arXiv: 1810.09235.

\bibitem{nitiss_formation_2020}
Edgars Nitiss, Tianyi Liu, Davide Grassani, Martin Pfeiffer, Tobias~J.
  Kippenberg, and Camille-Sophie Brès.
\newblock Formation {Rules} and {Dynamics} of {Photoinduced} chi(2) {Gratings}
  in {Silicon} {Nitride} {Waveguides}.
\newblock {\em ACS Photonics}, 7(1):147--153, January 2020.

\bibitem{duchesne_second_2011}
D.~Duchesne, K.~A. Rutkowska, M.~Volatier, F.~Légaré, S.~Delprat, M.~Chaker,
  D.~Modotto, A.~Locatelli, C.~De~Angelis, M.~Sorel, D.~N. Christodoulides,
  G.~Salamo, R.~Arès, V.~Aimez, and R.~Morandotti.
\newblock Second harmonic generation in {AlGaAs} photonic wires using low power
  continuous wave light.
\newblock {\em Optics Express}, 19(13):12408, June 2011.

\bibitem{poulvellarie_second-harmonic_2020}
Nicolas Poulvellarie, Utsav Dave, Koen Alexander, Charles Ciret, Maximilien
  Billet, Carlos Mas~Arabi, Fabrice Raineri, Sylvain Combrié, Alfredo
  De~Rossi, Gunther Roelkens, Simon-Pierre Gorza, Bart Kuyken, and François
  Leo.
\newblock Second-harmonic generation enabled by longitudinal electric-field
  components in photonic wire waveguides.
\newblock {\em Physical Review A}, 102(2):023521, August 2020.

\bibitem{ciret_influence_2020}
Charles Ciret, Koen Alexander, Nicolas Poulvellarie, Maximilien Billet, Carlos
  Mas~Arabi, Bart Kuyken, Simon-Pierre Gorza, and François Leo.
\newblock Influence of longitudinal mode components on second harmonic
  generation in {III}-{V}-on-insulator nanowires.
\newblock {\em Optics Express}, 28(21):31584, October 2020.

\bibitem{snyder_optical_1983}
Allan~Whitenack Snyder and John Love.
\newblock {\em Optical {Waveguide} {Theory}}.
\newblock Chapman and Hall, November 1983.

\bibitem{afshar_v_full_2009}
Shahraam Afshar~V. and Tanya~M. Monro.
\newblock A full vectorial model for pulse propagation in emerging waveguides
  with subwavelength structures part {I}: {Kerr} nonlinearity.
\newblock {\em Optics Express}, 17(4):2298, February 2009.

\bibitem{alloatti_second-order_2012}
L.~Alloatti, D.~Korn, C.~Weimann, C.~Koos, W.~Freude, and J.~Leuthold.
\newblock Second-order nonlinear silicon-organic hybrid waveguides.
\newblock {\em Optics Express}, 20(18):20506, August 2012.

\bibitem{kolesik_nonlinear_2004}
M.~Kolesik and J.~V. Moloney.
\newblock Nonlinear optical pulse propagation simulation: {From} {Maxwell}’s
  to unidirectional equations.
\newblock {\em Physical Review E}, 70(3):036604, September 2004.

\bibitem{ueno_second-order_1997}
Yoshiyasu Ueno, Vincent Ricci, and George~I Stegeman.
\newblock Second-order susceptibility of {Ga0}.{5In0}.{5P} crystals at 1.5 um
  and their feasibility for waveguide quasi-phase matching.
\newblock {\em Journal of the Optical Society of America B}, page~9, 1997.

\bibitem{driscoll_large_2009}
Jeffrey~B. Driscoll, Xiaoping Liu, Saam Yasseri, Iwei Hsieh, Jerry~I. Dadap,
  and Richard~M. Osgood.
\newblock Large longitudinal electric fields ({E}\_z) in silicon nanowire
  waveguides.
\newblock {\em Optics Express}, 17(4):2797, February 2009.

\bibitem{dave_nonlinear_2015}
Utsav~D. Dave, Bart Kuyken, François Leo, Simon-Pierre Gorza, Sylvain Combrie,
  Alfredo De~Rossi, Fabrice Raineri, and Gunther Roelkens.
\newblock Nonlinear properties of dispersion engineered {InGaP} photonic wire
  waveguides in the telecommunication wavelength range.
\newblock {\em Optics Express}, 23(4):4650, February 2015.

\bibitem{wang_second_2017}
Cheng Wang, Xiao Xiong, Nicolas Andrade, Vivek Venkataraman, Xi-Feng Ren,
  Guang-Can Guo, and Marko Lončar.
\newblock Second harmonic generation in nano-structured thin-film lithium
  niobate waveguides.
\newblock {\em Optics Express}, 25(6):6963, March 2017.

\bibitem{anthur_second_2020}
Aravind~P. Anthur, Zhang Haizhong, Yuriy Akimov, Ong Junrong, Dmitry
  Kalashnikov, Arseniy~I. Kuznetsov, and Leonid Krivitsky.
\newblock Second {Harmonic} {Generation} in {Gallium} {Phosphide}
  {Nano}-{Waveguides}.
\newblock {\em arXiv:2001.06142 [physics]}, January 2020.
\newblock arXiv: 2001.06142.

\bibitem{ottaviano_low-loss_2016}
Luisa Ottaviano, Minhao Pu, Elizaveta Semenova, and Kresten Yvind.
\newblock Low-loss high-confinement waveguides and microring resonators in
  {AlGaAs}-on-insulator.
\newblock {\em Optics Letters}, 41(17):3996, September 2016.

\bibitem{chang_ultra-efficient_2020}
Lin Chang, Weiqiang Xie, Haowen Shu, Qi-Fan Yang, Boqiang Shen, Andreas Boes,
  Jon~D. Peters, Warren Jin, Chao Xiang, Songtao Liu, Gregory Moille, Su-Peng
  Yu, Xingjun Wang, Kartik Srinivasan, Scott~B. Papp, Kerry Vahala, and John~E.
  Bowers.
\newblock Ultra-efficient frequency comb generation in {AlGaAs}-on-insulator
  microresonators.
\newblock {\em Nature Communications}, 11(1):1331, December 2020.

\bibitem{zaske_visible--telecom_2012}
Sebastian Zaske, Andreas Lenhard, Christian~A. Keßler, Jan Kettler, Christian
  Hepp, Carsten Arend, Roland Albrecht, Wolfgang-Michael Schulz, Michael
  Jetter, Peter Michler, and Christoph Becher.
\newblock Visible-to-{Telecom} {Quantum} {Frequency} {Conversion} of {Light}
  from a {Single} {Quantum} {Emitter}.
\newblock {\em Physical Review Letters}, 109(14):147404, October 2012.

\bibitem{diddams_direct_2000}
Scott~A. Diddams, David~J. Jones, Jun Ye, Steven~T. Cundiff, John~L. Hall,
  Jinendra~K. Ranka, Robert~S. Windeler, Ronald Holzwarth, Thomas Udem, and
  T.~W. Hänsch.
\newblock Direct {Link} between {Microwave} and {Optical} {Frequencies} with a
  300 {THz} {Femtosecond} {Laser} {Comb}.
\newblock {\em Physical Review Letters}, 84(22):5102--5105, May 2000.

\bibitem{okawachi_-chip_2020}
Yoshitomo Okawachi, Mengjie Yu, Boris Desiatov, Bok~Young Kim, Tobias Hansson,
  Marko Lončar, and Alexander~L. Gaeta.
\newblock On-chip self-referencing using integrated lithium niobate waveguides.
\newblock {\em arXiv:2003.11599 [physics]}, March 2020.
\newblock arXiv: 2003.11599.

\end{thebibliography}

\end{document}